\documentclass[preprint,secnumarabic,tightenlines,amssymb, amsmath,nobibnotes,
prd,nofootinbib,showpacs,]{revtex4}%

\usepackage[toc,page]{appendix}
\usepackage{dsfont}
\usepackage{graphicx}
\usepackage{dcolumn}
\usepackage{bm}
\usepackage{caption}
\usepackage{amssymb}
\usepackage{setspace}
\usepackage{amsmath}
\usepackage{amsfonts}
\usepackage{verbatim}
\usepackage{MnSymbol}
\usepackage{hyperref}
\usepackage{multirow}
\usepackage{color}%
\definecolor{darkgreen}{rgb}{0,0.35,0}

\definecolor{Rood}{rgb}{1, 0, 0}

\newcommand{\bfs}{\boldsymbol}
\newcommand{\be}{\begin{equation}}
\newcommand{\ee}{\end{equation}}
\newcommand{\beq}{\begin{eqnarray}}
\newcommand{\eeq}{\end{eqnarray}}
\newcommand{\beqs}{\begin{eqnarray*}}
\newcommand{\eeqs}{\end{eqnarray*}}

\begin{document}

\title{Casimir energy for N superconducting  cavities: \\
a model for the YBCO (BSCCO)}
\author{ A. Allocca$^{1,2}$, S. Avino$^{2,3}$, S. Balestrieri$^{1,4}$, E. Calloni$^{1,2}$, S. Caprara$^{ 5,6}$, M. Carpinelli$^{7,8}$,  \\
L. D'Onofrio$^{1,2}$, D. D'Urso$^{7,8}$, R. De Rosa$^{1,2}$, L. Errico$^{1,2}$, G. Gagliardi$^{2,3}$, 
M. Grilli$^{5,6}$,\\
 V. Mangano$^{5,6}$, M. Marsella$^{5,6}$, L. Naticchioni$^6$, A. Pasqualetti$^9$, G. P. Pepe$^{1,2}$, \\
M. Perciballi$^6$, L. Pesenti$^{10,11}$, P. Puppo$^6$, P. Rapagnani$^{5,6}$, F. Ricci$^{5,6}$, L. Rosa$^{1,2}$ \footnote{ luigi.rosa@unina.it}, 
\\
C. Rovelli$^{12,13,14}$, D. Rozza$^{7,8}$, P. Ruggi$^9$, N. L. Saini$^{5,6}$, V. Sequino$^{1,2}$, V. Sipala$^{7,8}$,
\\ 
 D. Stornaiuolo$^{1,2}$, F. Tafuri$^{1,2}$, A. Tagliacozzo$^{1,2}$, I. Tosta E Melo$^{7,8}$, L. Trozzo$^{2}$}

\affiliation
{
\begin{tabular}{l}
$^1$  Universit\`a di Napoli Federico II, 
Via Cintia Edificio 6, 80126 Napoli, Italy \\
$^2$ INFN Sezione di Napoli, 
Via Cintia Edificio 6, 80126 Napoli, Italy\\
$^3$ Centro Nazionale Ricerche-Istituto Nazionale di Ottica (CNR-INO),  \\
~~SS Napoli, Via Campi Flegrei 34, 80078 Pozzuoli (NA), Italy \\
$^4$ CNR-ISASI, Via Pietro Castellino 111, 80131 Napoli, Italy.\\
$^5$Universit\`a di Roma ``La Sapienza'', P.le A. Moro 2, I-00185, Roma, Italy\\
$^6$ INFN Sezione di Roma, P.le A. Moro 2, I-00185 Roma, Italy\\
$^7$Universit\`a degli studi di Sassari, I-07100 Sassari, Italy\\
$^8$ Laboratori Nazionali del Sud, I-95125 Catania, Italy\\
$^9$ European Gravitational Observatory (EGO), I-56021 Cascina (Pi), Italy\\
$^{10}$ Universit\`a degli Studi di Milano Bicocca, I- 20126 Milano, Italy\\
$^{11}$ INFN - sezione di Milano-Bicocca, I - 20126 Milano, Italy\\ 
$^{12}$Centre de Physique Theorique Campus of Luminy - Case 907, F-13288 Marseille, France.\\
$^{13}$ Aix Marseille Universit\'e CNRS, CPT, UMR 7332, 13288,  Marseille, France. \\
$^{14}$Universit\'e de Toulon, CNRS, CPT, UMR 7332, 83957 La Garde, France\\
\\
\end{tabular}
}

\begin{abstract}
In this paper we study the Casimir energy of a sample made by $N$ cavities, with $N\gg1$, across the transition from the metallic 
to the superconducting phase of the constituting plates. After having characterised the energy for the configuration in which 
the layers constituting the cavities are made by dielectric and for the configuration in which the layers are made by plasma sheets, 
we concentrate our analysis on the latter. It represents the final step towards the macroscopical characterisation of 
a ``multi cavity'' (with $N$ large) necessary to fully understand the behaviour of the Casimir energy of a YBCO (or a BSCCO) sample 
across the transition.

Our analysis is especially useful to the Archimedes experiment, aimed at measuring the interaction of the electromagnetic 
vacuum energy  with a gravitational field. To this purpose, we aim at modulating the Casimir energy of a layered structure, the multi cavity,  
by inducing a transition from the metallic to the superconducting phase.
After having characterised the Casimir energy of such a structure for both the metallic and the superconducting phase, we give an 
estimate of the modulation of the energy across the transition. 
\end{abstract}

\pacs{12.20.Ds, 12.20.-m,  74.72.-h }


\maketitle

\section{Introduction}

The principal goal of the  Archimedes experiment \cite{erico_2014} is to measure the coupling of the vacuum fluctuations
of Quantum Electrodynamics (QED) to the gravitational field of the Earth. The coupling is obtained, as usual in Quantum Field Theory 
in Curved spacetime \cite{bimo_2006,fulling_2007,bimo_2008,espo_2008}, assuming the Einstein tensor  to be proportional to the 
expectation value of the regularized and renormalized energy-momentum tensor of matter fields, in particular, for the Archimedes 
experiment, of the electromagnetic field. The idea is to weigh the vacuum energy stored in a rigid Casimir cavity \cite{cas53}, made by parallel 
conducting plates, by modulating the reflectivity of the plates upon inducing a transition from the metallic to the superconducting 
state \cite{erico_2014}. The ``modulation factor'' is defined as $\eta=\frac{\Delta E}{E}$ were $\Delta E$ is the 
difference of Casimir energy in the normal and in the superconducting state.

In Ref.\,\cite{erico_2014} it was shown that, in order to measure such an effect, $\eta$ must be of the order $\eta\sim 10^{-5}$ and 
that, to this purpose, a multi cavity, obtained by superimposing many cavities must be used. This structure is natural in the case of 
crystals of type-II superconductors, particularly cuprates, being composed by Cu-O planes, that undergo the superconducting transition, 
separated by nonconducting planes. A crucial aspect to be tested is  the behavior of the Casimir energy \cite{cas53} for a 
multi cavity when the layers undergo the phase transition from the metallic to the superconducting phase. In a previous paper 
\cite{Rosa:2017tfo} a careful study for such a type of structure has been carried out for a sample made by up to three ``relatively 
thick'' (of the order of ten nanometer) dielectric layers. In the present paper we extend the analysis to any number of cavities 
for both situations: layers consisting of ``thick'' dielectric slabs and layers consisting of ``thin'' plasma sheets.
 
Indeed, in Ref.\,\cite{kempf}, considering a cavity based on a high-$T_{\mathrm c}$ layered superconductor, a factor as high as 
$\eta=4\times 10^{-4}$ has been estimated (for flat plasma sheets at zero temperature and no conduction in the normal state, so 
that $\Delta E$ corresponds to the energy of the ideal cavity, and charge density $n = 10^{14}$\,cm$^{-2}$). 
The Archimedes sensitivity is expected to be capable of assessing the interaction of gravity and vacuum energy also for values 
lower than $\eta=4\times 10^{-4}$, up to $1/100$ of this value \cite{erico_2014}. It is then crucial to 
understand the level of modulation achievable with layered superconducting structures. This is the scope of the present paper.
 
Considering in particular the multi cavity, the general assumption adopted so far has been that the Casimir energy obtained by
overlapping many cavities is the sum of the energies of each individual cavity. This is true if the {\em distances} between 
neighboring cavities are large (in the sense that the thickness of each metallic layer separating the various cavities is very 
large with respect to the penetration depth of the radiation field). Of course, this is no longer true if the thickness of these 
metallic inter-cavity layers gets thinner and thinner. 

Sec.\,II studies the Casimir energy of a multilayered cavity, assuming either dielectric or plasma sheet matching conditions at 
each interface between the layers. In Sec.\,III, numerical calculations are carried out and an analytic model capable of describing 
the Casimir energy at finite temperature is given. Finally, in Sec.\,IV, a possible model for describing the variation (and the 
modulation) of the Casimir energy across the transition is introduced. Our concluding remarks are found in Sec.\,V. 

\section{The Casimir energy of $N$ coupled cavities}
In this section we deduce the Casimir Energy of $N$ coupled cavities, even though in the present paper we are interested in 
applying our results to plasma sheets, we will discuss the case of dielectrics first and then recover the plasma sheets results as 
a suitable limiting case.

In the following, referring to Fig.\,1, $d_i$ is the distance of the $i$-th cavity from the 
$(i-1)$-th, (thickness of the $i$-th cavity), within the slabs $1,3$ and $5$ there is vacuum while the regions $0,2,4$ and $6$ 
are dielectric. The thickness of the regions $0$ and $6$ is assumed to be infinite.

\begin{center}
 \begin{figure}[ht]
     \centering
    \includegraphics[width = .6\textwidth]{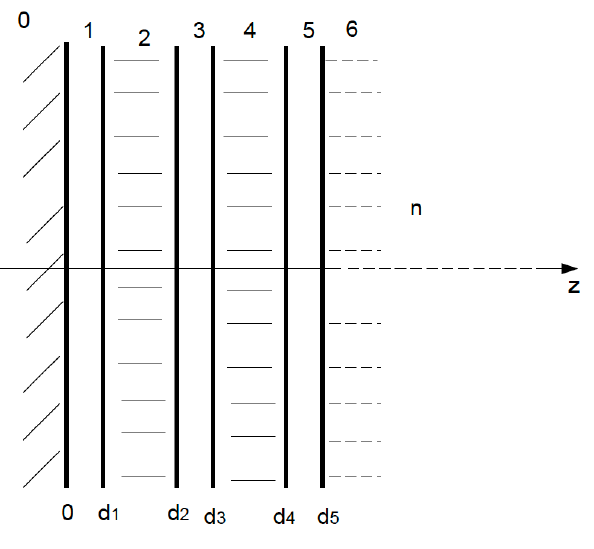}
   \caption{A $N$ layer cavity. For the dielectric case in the 0, 2, 4, ..., and all even-numbered regions there is a dielectric 
   and in the 1, 3, 5, ..., and all odd-numbered regions there is vacuum. $d_i$ is the thickness of the $i$-th slab. In the case of 
   plasma-sheet there is vacuum everywhere and the layer are simple interfaces (of zero thickness) at $z=d_1$, $d_2$, ...
}
\label{fig:ncav}
  \end{figure}
\end{center}

The general expression for the Casimir energy (per unit area), at finite temperature, will be written in the usual manner 
\cite{milton_01,bord09,bord01} 
\begin{equation} 
E=\, {k_B
\,T}\sum_{l=0}^{~\infty\,~\prime}\int \frac{ \rm d {\bf k_{\bot}}}{(2
\pi)^2} \,\left[\log{\Delta^{{\rm TE}}(\zeta_l)}+\log
{\Delta^{{\rm TM}}(\zeta_l)}\right]
\label{fint1}
\end{equation}
where the $\Delta$ are the so called generating functions (in the following we will omit the subscript $\rm TM(TE)$ if no ambiguity is 
generated), $\zeta_l=2\pi l k_B T$ are the Matsubara frequencies, $k_B$ is the Boltzmann constant, 
$l={0,1,2,\ldots}$, and the superscript $^\prime$ on the sum means that the zero mode must be multiplied by a factor 
$\frac{1}{2}$. The generating functions are obtained by computing the determinant of the most general boundary conditions at 
each singular layer  located at  $d_0, d_0+d_1,d_0+d_1+d_2$...etc.  (see Fig.\,\ref{fig:ncav}; see also the appendix) \cite{jack98}.

For the sake of clarity, we only give here the general argument about the procedure for obtaining the generating 
functions, referring the reader to the appendix  for the complete computation.
In  the appendix we show that the $\Delta$ functions can be written in terms of  a sort of generalised reflection coefficients:  
\begin{eqnarray*}
R_{\rm TM}^{i,j} &=& \frac{\epsilon_j(\mathrm i\zeta_l)K_i-\epsilon_i(\mathrm i\zeta_l)K_j - 
{2\frac{\Omega}{\zeta_l^2}} K_iK_j}{\epsilon_j(\mathrm i\zeta_l)K_i+\epsilon_i(\mathrm i\zeta_l)K_j + 
{2\frac{\Omega}{\zeta_l^2}} K_iK_j},~~~~~R_{\rm TE}^{i,j} = \frac{K_i-K_j +2 \Omega}{K_i+K_j +2 \Omega}, \\
 \\
S_{\rm TM}^{i,j} &=&\frac{\epsilon_j(\mathrm i\zeta_l)K_i-\epsilon_i(\mathrm i\zeta_l)K_j + 
{2\frac{\Omega}{\zeta_l^2}} K_iK_j}{\epsilon_j(\mathrm i\zeta_l)K_i+\epsilon_i(\mathrm i\zeta_l)K_j + 
{2\frac{\Omega}{\zeta_l^2}} K_iK_j},~~~~~S_{\rm TE}^{i,j} = \frac{K_i-K_j -2 \Omega}{K_i+K_j +2 \Omega}, \\ \\
T_{\rm TM}^{i,j} &=&\frac{\epsilon_j(\mathrm i\zeta_l)K_i+\epsilon_i(\mathrm i\zeta_l)K_j - 
{2\frac{\Omega}{\zeta_l^2}} K_iK_j}{\epsilon_j(\mathrm i\zeta_l)K_i+\epsilon_i(\mathrm i\zeta_l)K_j + {2\frac{\Omega}{\zeta_l^2}} K_iK_j},
~~~~~T_{\rm TE}^{i,j} = \frac{K_i+K_j -2 \Omega}{K_i+K_j +2 \Omega}, 
\end{eqnarray*}
where $K_i=\sqrt{k_\perp^2+\epsilon_i(\mathrm i\zeta_l) \zeta^2_l}$, $k_\perp=(k_x,k_y)$, $\Omega=\frac{\mu_0 n_{2D} q^{*2}}{m^{*} }$,
$\mu_0$ is the magnetic permeability of vacuum, $n_{2D}$ is the  two dimensional carrier density in the layer, and $q^*$ and $m^*$, 
respectively, their charge and mass. The standard dielectric boundary conditions (dbc) will be recovered by imposing  $\Omega=0$ and 
the plasma sheet boundary conditions (psbc) by requiring $\epsilon_i(\mathrm i\zeta_l)=1,~~\forall i$ (in this case, $K_i=K_j$).

After introducing the auxiliary functions
\begin{eqnarray*}
{E}^{ijk}  &=&  {\mathrm e^{-2 d_j K_j} \text{S}^{j,k} \text{R}^{i,j}}+1, \\
F^{ijk}  &=&{\mathrm e^{-2 d_j K_j} \text{R}^{i,j}\text{T}^{j,k}}+\text{R}^{j,k}, \\
G^{ijk} &=& {\mathrm e^{-2 d_j K_j} \text{S}^{j,k} \text{T}^{i,j}}+{\text{S}^{i,j}}, \\
H^{ijk}   &=&{\text{S}^{i,j} \text{R}^{j,k}}+{\mathrm e^{-2 d_j K_j} \text{T}^{i,j} \text{T}^{j,k}}
\end{eqnarray*}
and (henceforth, we will assume all the cavities to be equal and consider only the indices $\{ijk\}= \{012\}$)
\begin{eqnarray*}
I_1 &=& E^{012};~~I_2=F^{012}\mathrm e^{-2 d_2  K_2}G^{012},\\
I_n &=&F^{012}\mathrm e^{-2 d_2  K_2}\left(H^{012}\mathrm e^{-2 d_2  K_2}\right)^{n-2}G^{012}, ~~~\mbox{      for $n\geq3$},
\end{eqnarray*}
we can proceed to compute the generating functions. 

\subsection{The dielectric case}

Let us consider Casimir cavities made of dielectric layers (of thickness $d_i$). To obtain the general expression for the $\Delta$ 
functions we can proceed inductively (a very detailed discussion up to three cavities can be found in Ref.\,\cite{Rosa:2017tfo}). For  
the cavity characterised  by the numbers $(012)$ in Fig.\,1, with $\epsilon_{0}=\epsilon_{2}$, the generating function,  
for $\rm TM$ and $\rm TE$ modes, respectively, is obtained in the usual manner \cite{bord09,Rosa:2017tfo} (see appendix).
After regularization, i.e., setting to zero the Casimir energy when the two cavities are infinitely far away, the result 
can be written as $\Delta_1=E^{012}=I_1$. Let us now consider two cavities [$(012)$, $(234)$ in Fig.\,1]. 
In this case, the generating function is the determinant of the $8\times8$ matrix made by the first rows and columns of the matrix 
given in the appendix  \cite{Rosa:2017tfo}. It can be written as a $2\times2$ block matrix, thus \cite{Silvester_2000,Powell_2011}
\[
\Delta=\mathrm{det}
\left(
\begin{array}{cc}
A&  B    \\
 C &  D    \\   
\end{array}
\right)=\mathrm{det}(A)\,\mathrm{det}(\mathds{1}-A^{-1}BD^{-1}C)\,\mathrm{det}(D),
\]
where $\{A,B,C,D\}$ are $4\times4$ matrices, with $\mathrm{det}(A)=\mathrm{det}(D)=\Delta_1$.

When the two cavities are infinitely far away from each other ($d_2\rightarrow\infty$), $C=0$, 
$\Delta=\mathrm{det}(A)\,\mathrm{det}(D)=:\Delta_2$ and the Casimir energy will be simply the sum of the energies of the two cavities,
$\log{(\Delta_2)}=\log{(\Delta_1^2)}=2 \log{(\Delta_1)}$. 
When they are brought at a distance $d_2$ from each other, in addition to the previous energy, there is the interaction energy accounted 
for by the term $\mathrm{det}(\mathds{1}-A^{-1}BD^{-1}C)$.
In this case  $\Delta_2=\mathrm{det}(A)\,\mathrm{det}(D)\,\mathrm{det}(\mathds{1}-A^{-1}BD^{-1}C)$ and, after regularization,
it can be written (see appendix) as $\Delta_2=:I_1^2+I_2$, so that the corresponding Casimir energy depends on 
$\log{\Delta_2}=\log{(I_1^2+I_2)}=\log{(I_1^2)}+\log{(1+I_2/I_1^2)}$. The first term is simply the sum of the energies of the two 
cavities taken independently, the second term is the interaction energy between the two \cite{Rosa:2017tfo}. Therefore  we can always reduce 
ourselves to the computation of determinants of products of $4\times4$ matrix. The interaction in the case of $n\geq 3$ cavities is 
accounted for by the term $I_n$.

In this manner, using the inductive principle, it is not difficult to convince oneself that the generic $\Delta_N$ functions 
for the case of $N$ dielectric cavities can be obtained in the following manner (a sort of Feynman diagram for the generating functions):
let us define $\{k_1,k_2,\ldots,k_J\}$ to be the  $J$-th integer partition of $N$ and $Q_J$ its multiplicity (the number of combinations 
that contain the same type of $I_k$ but in a different position) then
$$
\Delta_N=\sum_{J} Q_J\left(I_{k_1} I_{k_2}\ldots I_{k_J}\right).
$$
So, for example,
\begin{eqnarray*}
\Delta_{1} &=& I_1, \\
\Delta_{2} &=& (I_1)^2+I_2 , \\
\Delta_{3} &=& (I_1)^3+I_1I_2+I_2I_1+I_3= (I_1)^3+2I_1I_2+I_3 ,\\
\Delta_{4} &=& I_4+I_1I_3+I_3I_1+I_2^2+I_1I_1I_2+I_1I_2I_1+I_2I_1I_1+I_1^4 \nonumber \\
&=& I_4+2I_1I_3+I_2^2+3I_1^2I_2+I_1^4 ,
\end{eqnarray*}
and, e.g.,
\begin{eqnarray*}
\Delta_{10} &=& I_{1}^{10} + 9 I_{1}^8 I_{2} + 28 I_{1}^6 I_{2}^2 + 
 35 I_{1}^4 I_{2}^3 + 15 I_{1}^2 I_{2}^4 + I_{2}^5 + 
 8 I_{1}^7 I_{3} + 42 I_{1}^5 I_{2} I_{3} + 
 60 I_{1}^3 I_{2}^2 I_{3} + 20 I_{1} I_{2}^3 I_{3} +  \\ 
 &&  15 I_{1}^4 I_{3}^2 +30 I_{1}^2 I_{2} I_{3}^2 + 6 I_{2}^2 I_{3}^2 + 
 4 I_{1} I_{3}^3 + 7 I_{1}^6 I_{4} + 30 I_{1}^4 I_{2} I_{4} + 
 30 I_{1}^2 I_{2}^2 I_{4} + 4 I_{2}^3 I_{4} + 
 20 I_{1}^3 I_{3} I_{4} + \\ 
 && 24 I_{1} I_{2} I_{3} I_{4} + 
 3 I_{3}^2 I_{4}+ 6 I_{1}^2 I_{4}^2 + 3 I_{2} I_{4}^2 + 
 6 I_{1}^5 I_{5} + 20 I_{1}^3 I_{2} I_{5} + 12 I_{1} I_{2}^2 I_{5} + 
 12 I_{1}^2 I_{3} I_{5} + 6 I_{2} I_{3} I_{5} +   \\ 
 &&6 I_{1} I_{4} I_{5} + 
I_{5}^2 + 5 I_{1}^4 I_{6} +12 I_{1}^2 I_{2} I_{6} + 
 3 I_{2}^2 I_{6} + 6 I_{1} I_{3} I_{6} + 2 I_{4} I_{6} + 
 4 I_{1}^3 I_{7} + 6 I_{1} I_{2} I_{7} + \\
 &&2 I_{3} I_{7} + 
 3 I_{1}^2 I_{8} + 2 I_{2} I_{8} + 2 I_{1} I_{9} + I_{10},
\end{eqnarray*} 
for ten cavities. 

\subsection{The Plasma Sheets case}
These formulae can be extended to the case in which the layers are characterised as plasma sheets.
For example, the two dielectric cavities $(012)$ and $(234)$ can describe three plasma-sheet cavities, $(012),~(123),~(234)$, by 
imposing $\epsilon_i=1$, and $\Omega\neq 0$. In other words, two dielectric cavities needs four layers located at $0,\,d_1,\,d_1+d_2,\,d_1+d_2+d_3$ but the same four layers correspond to three cavities having plasma sheet as boundaries.
Consequently $N_{ps}$ (odd) plasma sheets can be obtained by $n=\frac{N_{ps}+1}{2}$ standard dielectrics by simply imposing  
$\epsilon_i(\mathrm i\zeta)=1$, and the extension of the previous formulae to the case of an odd number of plasma sheets is 
straightforward.

The case of an even number of plasma sheets is more involved. It can be obtained starting with $N_{ps}+1$ ($N_{ps}$ even) cavities and 
moving the last layer to infinity. From the mathematical point of view, this procedure corresponds to introducing a term $I'_n$  
(which describes the interaction of the last interface with all the others), defined like as
\be
I'_1=1;~~~~ I'_n=\lim_{G\rightarrow G'}I_n,~~\mbox{if}~~n\geq2 ; \mbox{    with  }G'^{~ij}=S^{ij}.
\ee
In this manner, we have for two and four plasma sheet (please note that it is necessary to perform the limiting procedure first 
and then to group together the various terms)
\begin{eqnarray}
\Delta_2^{ps} &=& \lim_{G\rightarrow G'}\Delta_{2+1}^{ps} =\lim_{G\rightarrow G'}\Delta_2=\lim_{G\rightarrow G'}
\left[ (I_1)^2+I_2 \right] =I_1I_1'+I_2'=I_1+I_2'; \\
\Delta_4^{ps} &=& \lim_{G\rightarrow G'}\Delta_{4+1}^{ps} =\lim_{G\rightarrow G'}\Delta_3=\lim_{G\rightarrow G'}
\left[ (I_1)^3+I_1I_2+I_2I_1+I_3 \right] \nonumber \\
&=&I_1I_1I'_1+I_1I'_2+I_2I'_1+I'_3=I_1^2+I_2+I_1I'_2+I'_3.
\end{eqnarray}
The fact that only one term at a time takes the prime corresponds to the fact that the last cavity only must be sent to infinity.

\section{Numerical results}
We are now in the position to discuss the dependence of the Casimir Energy of 
a $N$-cavity made of $N-$plasma sheets.

We start by considering the variation of the Casimir energy as a function of the number of cavities for fixed thickness $d_i=2$\,nm and 
 $\Omega=\frac{\mu_0 n_{2D} q^{*2}}{m^{*} } =49593.3$\,m$^{-1}$ (see Refs.\,\cite{orlando_2018,harshman_1992}).
We get $\frac{E[1]}{A}=-0.000197\,{J}{m^{-2}}$ and, for the ratio $\frac{E[N]}{N E[1]}$  
between the Casimir Energy of $N$ cavities $E[N]$, and the product $N E[1]$ between the number of cavities and the energy of a 
single cavity $E[1]$, we find the values quoted in the following table. (NB We underline the fact that the contribution of TE modes results various order of magnitude less than the one from TM modes. For this reason, in the following, it will be simply omitted)

\begin{table}[ht]
\begin{center} 
\[\begin{array}{|c|c|}\hline
N &\frac{1}{N}\frac{E[N]}{E[1]}\\
& \\ \hline
 1 & 1.0000 \\ \hline
 2 & 1.0125 \\ \hline
 3 & 1.0181\\ \hline 
  4 & 1.0212\\ \hline 
 5 & 1.0232 \\ \hline 
 6 & 1.0246 \\ \hline
 7 & 1.0256 \\ \hline
 8 & 1.0263 \\ \hline
 9 & 1.0269 \\ \hline
 10 & 1.0274 \\ \hline
 11 & 1.0278 \\ \hline
 13 & 1.0284 \\ \hline
 15 & 1.0288 \\ \hline
 17 & 1.0292 \\ \hline
 19 & 1.0294 \\ \hline
\end{array}\]
\caption{The ratio $\frac{E[N]}{N E[1]}$ as a function of the number of cavities}
\end{center}
\label{default}
\end{table}

\begin{center}
 \begin{figure}[ht]
     \centering
    \includegraphics[width = .6\textwidth]{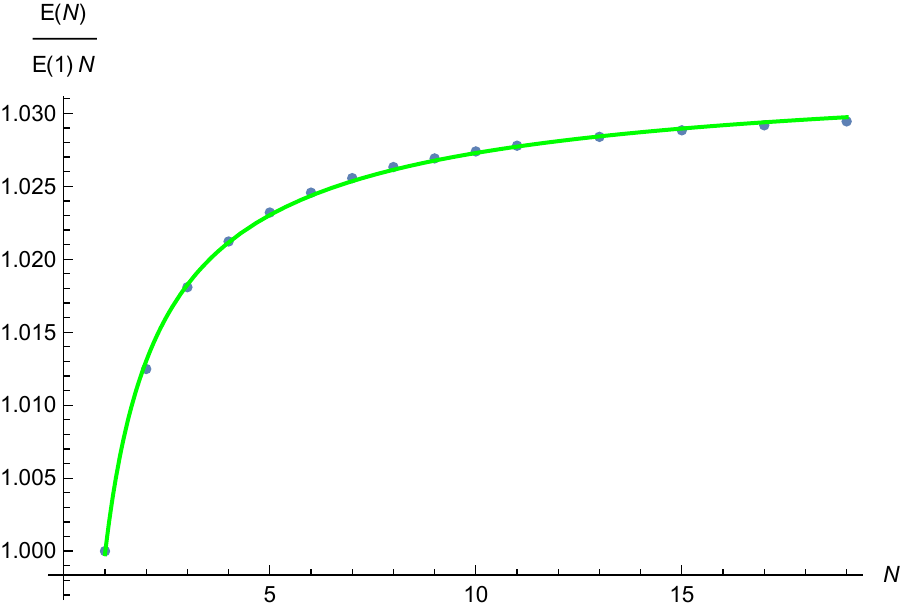}
   \caption{ The exact numerical result (dots) and the fitted results given by Eq.\,(\ref{eq:fit_1}) (green line) of the function 
   $ E[N]/(N E[1])$ for $d=2$\,nm.}
   \label{fig:asymcas}
  \end{figure}
\end{center}

The best fit is given by
\begin{equation}
\frac{1}{N}\frac{E[N]}{E[1]}=1.034\, -\frac{0.034}{N^{0.71}},
\label{eq:fit_1}
\end{equation}
that gives a clear indication of the presence of an asymptote for $N\rightarrow\infty$.
In Fig.\,\ref{fig:asymcas} a comparison between the exact numerical result and the analytical fitted behaviour up to 
$N=19$ [Eq.\,(\ref{eq:fit_1})] is shown.

Thus, we obtained an asymptotic expression for the Casimir energy for large $N$,
\begin{equation}
E[N]\simeq(1.034\, E[1]) N
\end{equation}
and deduced that the coupling of the various cavities resulted in an increase of the Casimir energy of $3.4\%$. This result 
is very different from the result in the case of dielectric layers \cite{Rosa:2017tfo}, in which a very strong coupling between 
different layers was found, mainly due to the zero Matsubara mode.
%

Remembering that \cite{bordag_2006} $E[1]=5 \times 10^{-3}\hbar c \sqrt{\frac{\Omega}{d^5}}$, with 
$\Omega=\frac{ n_{2D}q^2}{2 m^*\epsilon_0 c^2}$,
it seems reasonable to assume  $E[N]/A=-(1.034 \,N)K\hbar c \frac{\Omega^\alpha}{d^\beta}$, estimating the best values for 
${K,\alpha,\beta}$. We found $K=5.0\times 10^{-3}$, $\beta=2.4998$ and $\alpha=0.4998$,
 in perfect agreement with Ref.\,\cite{bordag_2006}.  In conclusion  the Casimir energy (per unit surface) of $N$ plasma sheet at fixed temperature 
$T=94$\,K is given by
\begin{equation}
E[N]/A=-(1.034 K \hbar c \frac{\sqrt{\Omega}}{d^{5/2}}) N \left({J}{m^{-2}}\right)\label{eq:fitteden}
\end{equation}

\begin{center}
 \begin{figure}[ht]
     \centering
    \includegraphics[width = .6\textwidth]{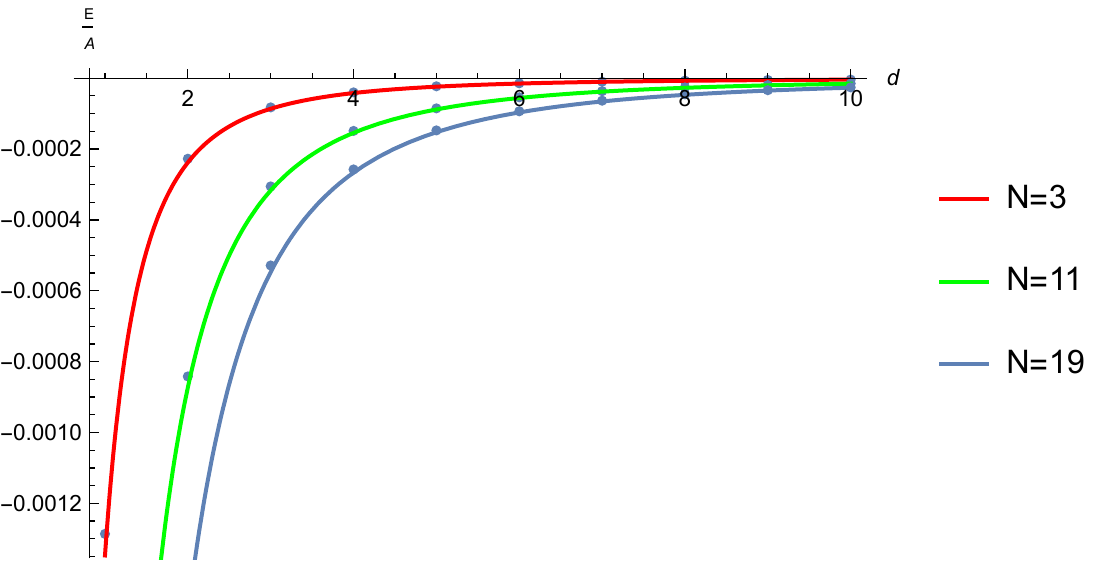}
   \caption{A comparison between exact numerical values of the Casimir energy (dots) and the approximated formula, Eq.\,(\ref{eq:fitteden}) (lines),
   with $d$ expressed in nm and $E/A$ in $J/m^2$.
   }
   \label{fig:vard}
  \end{figure}
\end{center}

In Fig.\,\ref{fig:vard} the exact values (dots) of the Casimir Energy as a function of $d$ are shown for $N=\{3,11,19\}$ and compared
with the fitting formula, Eq.\,(\ref{eq:fitteden}).

\begin{center}
 \begin{figure}[ht]
     \centering
    \includegraphics[width = .6\textwidth]{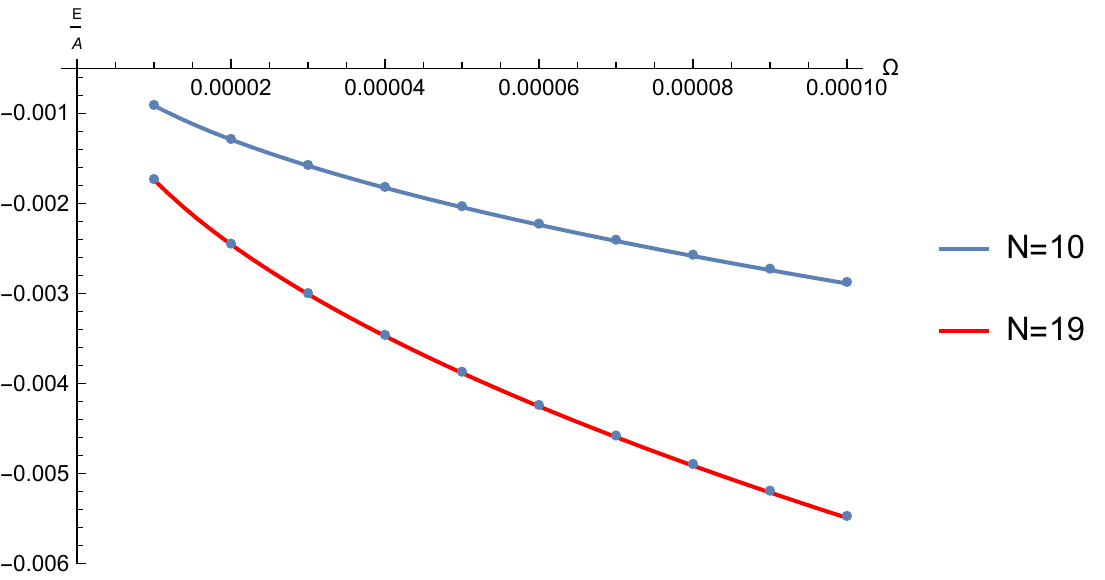}
   \caption{ A comparison between exact values of the Casimir energy (dots) and the approximated formula (lines), for $d=2$\,nm, with $\Omega$ expressed in nm$^{-1}$ $E/A$ in $J/m^2$.}
    \label{fig:varom_1}
  \end{figure}
\end{center}

In Fig.\,\ref{fig:varom_1} the Casimir Energy is plotted as a function of $\Omega$ for $N=\{10,19\}$.

In conclusion, a good approximation for the Casimir Energy (at fixed temperature) for $N$ plasma-sheet cavities can be written as
\begin{equation}
\frac{E[N]}{A}=-\left(1.034\, K \hbar c \frac{\sqrt{\Omega}}{d^{5/2}}\right) N=\left(-1.63 \times 10^{-28}(J m)\right)\left(N\frac{\sqrt{\Omega}}{d^{5/2}}(m^{-3})\right),
\end{equation}
with ${E[N]}/{A}$ measured in $J m^{-2}$.

Based on the above formulae, in the following section we give an estimate for the variation of the Casimir energy across the 
metal-superconductor transition. 
 
\section{The variation of the Casimir energy in the YBCO}

\begin{center}
 \begin{figure}[ht]
     \centering
    \includegraphics[width = .6\textwidth]{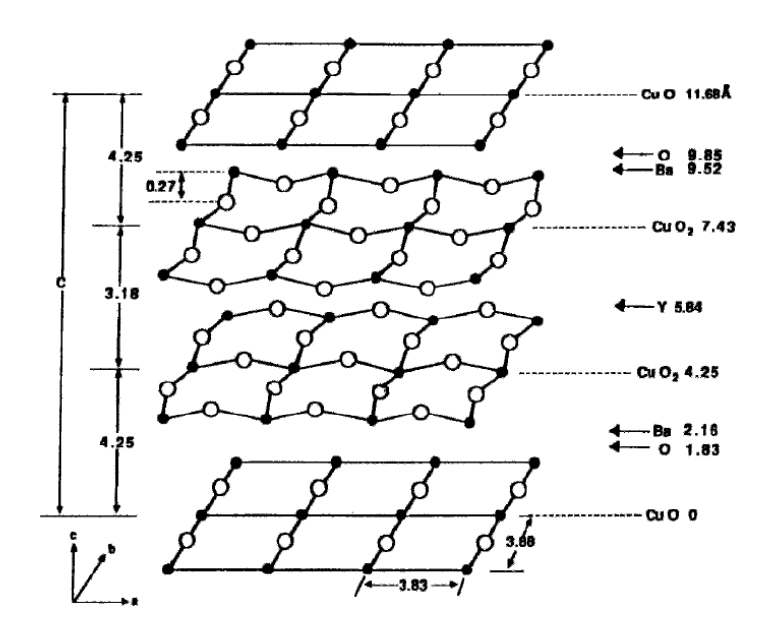}
   \caption{Typical layered strucutre of $YBa_2Cu_3O_7$ \cite{poole}.}
    \label{fig:ybcolayer_1}
  \end{figure}
\end{center}

The typical structure of a YBCO cell is represented in Fig.\,\ref{fig:ybcolayer_1}, in which
$\delta=4.25 \rm{\AA}$ is the thickness of our plasma sheet and $d=3.18\rm \AA$ is the distance between the layers.

By observing that \cite{orlando_2018} $\Omega(0)=\frac{\delta}{2 \lambda_{ab}^2(0)}$, at $T=0$\,K,
we can write for the Casimir energy of one cavity in the superconducting state as 
\beq
\frac{E (0)}{A}=-1.63\times 10^{-28}{\sqrt{\frac{\delta}{2d^5}}\frac{1}{\lambda_{ab}(0)}}.\eeq 
Using the BCS relation $\lambda(T)=\frac{\lambda(0)}{\sqrt{1-(T/T_{\mathrm c})^{4/3}}}$ \cite{poole}, corresponding to the case of 
$d$-wave pairing, as it is suitable for cuprates, for $T<T_{\mathrm c}$ and for one cavity we get
\beq
\frac{E (T)}{A} &=&-1.63\times 10^{-28}{\sqrt{\frac{\delta}{2d^5}}\frac{1}{\lambda_{ab}(T)}} =-1.63
\times 10^{-28}{\sqrt{\frac{\delta}{2d^5}}\frac{\sqrt{1-(T/T_{\mathrm c})^{4/3}}}{\lambda_{ab}(0)}}.
\eeq
Thus, using for YBa$_2$Cu$_3$O$_7$ \cite{harshman_1992}, $T_{\mathrm c}=92$\,K, $\lambda_{ab}(0)=1415\,\rm\AA$, $d=3.36\,\rm \AA$, 
and $\delta=5.84\,\rm\AA$, we have
\beq
\frac{E (90)}{A}
&=&-9.51\times10^{-3}\sqrt{1-(90/92)^{4/3}} 
=-0.001616~~ J m^{-2}.
\eeq
For the normal phase, $T>T_{\mathrm c}$, we will use the data (and formulae) of Ref.\,\cite{palenskis_2015}, where 
particular attention is paid to the treatment of semiconductors with degenerate electron gases.
At $T=100$\,K, we have
$n_{3D}=3.1\times 10^{25}$\,m$^{-3}$, which implies $n_{2D}=\delta n_{3D}= (5.84\times 10^{-10})
(3.1\times 10^{25})=1.810\times 10^{16}\,m^{-2}$, so that at $T=94$\,K, we have 
$n_{2D}= 1.317\times 10^{16} \frac{94}{100}=1.702 \times 10^{16}$\,m$^{-2}$. 
Consequently, $\Omega=\frac{\mu_0 n_{2D} e^2}{2m^*}=300.505$\,m$^{-1}$, and
\[
\frac{ E (94)}{A}=-1.63\times10^{-28}\sqrt{\frac{\Omega}{d^5}}=-0.001365~~ J m^{-2}.
\]
Thus, 
\[
\frac{\Delta E}{A}=\frac{E (94)-E (90)}{A}=0.000251~~ J m^{-2}.
\]
 
\begin{center}
 \begin{figure}[ht]
     \centering
    \includegraphics[width = .5\textwidth]{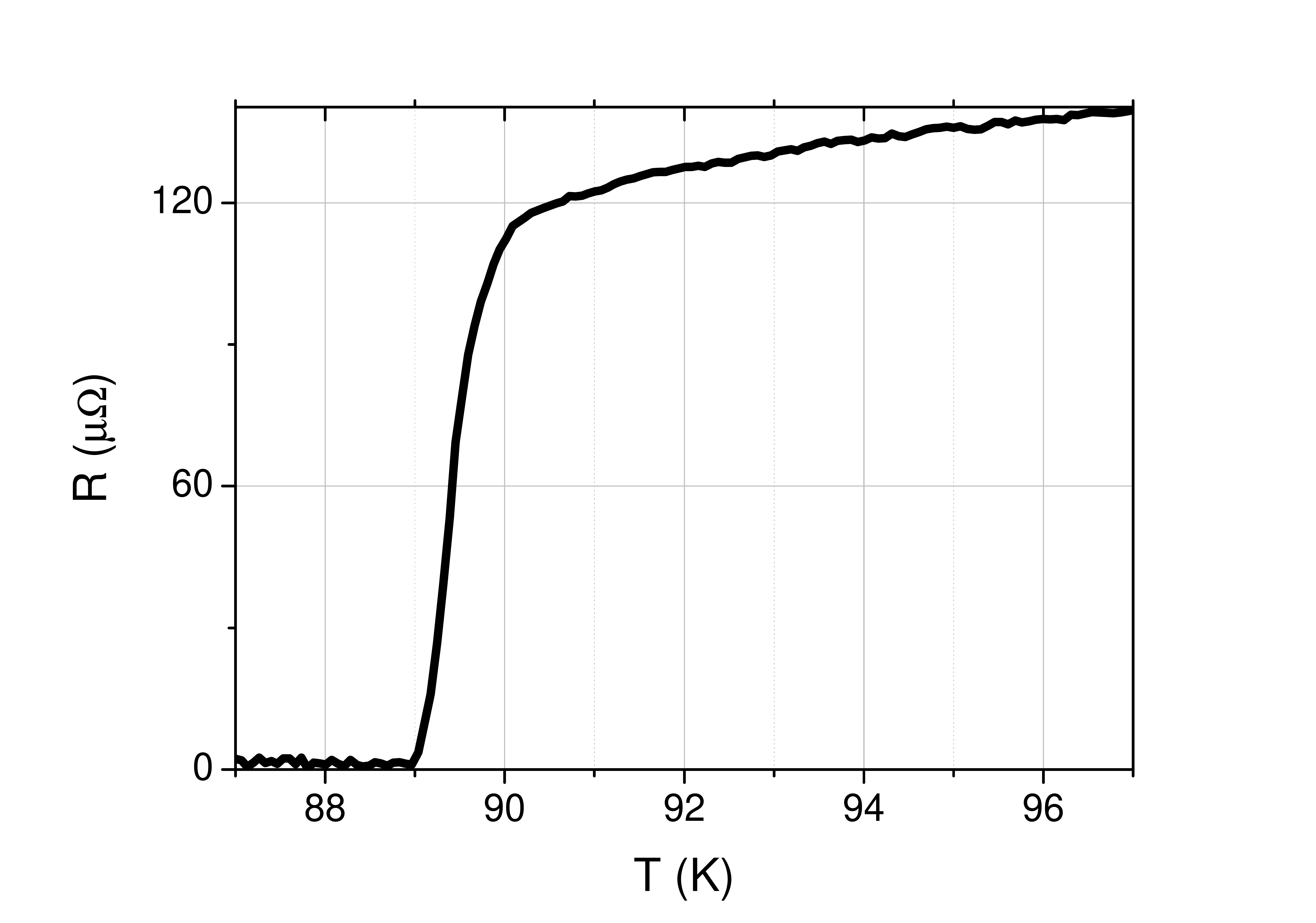}
   \caption{The transition of the sample of YBCO we are using.}
    \label{fig:ybcotransition_1}
  \end{figure}
\end{center}

As revealed by an inspection of the table in Ref.\,\cite{harshman_1992},
it is clear that the previous results depend in a crucial way on the sample of YBCO used.
A typical Resistance vs. Temperature curve of the YBCO crystals we are planning to use in the Archimedes experiment is reported in 
Fig.\,\ref{fig:ybcotransition_1}. Our reference values are $T_{\rm c}=89~K$ and $\lambda_{ab}(0)=1030\,\rm\AA$ 
\cite{Collignon_2017}, thus, assuming all the other parameters unaltered, we have
$\frac{E (87)}{A}=-0.002258~ Jm^{-2}$ and, in the normal phase, $T=91$\,K,
$\frac{E (91)}{A}=-0.001343~ Jm^{-2}$  so that $\frac{\Delta E}{A}=\frac{E (91)-E (87)}{A}=0.0009142~J m^{-2}$.

\section{Conclusions}
We have proposed a model for computing the variation of the Casimir energy of a YBCO sample across the metal-superconductor 
transition. 

We have constructed a powerful procedure to compute the renormalised Casimir energy both in the case of cavities made of a large 
number of thick dielectric layers and in the case of cavities made by a large number of thin plasma sheet layers. 

Our main assumption is that the last case can be used to describe the Casimir energy in YBCO and, more generally in cuprates (BSCCO), 
because of their natural built-in layered structure, both in the normal and in the superconducting phase.

We suggested a possible way of characterising the variation of the Casimir energy at the metal-superconductor transition, giving 
a numerical estimate for the specific YBCO sample that we are using in the Archimedes experiment.

The computed value for the relative variation of the Casimir energy, for one cavity, is  
$\frac{\Delta E}{E} =\frac{0.00091}{0.0013}=0.7$, which is quite reassuring for the Archimedes experiment.

It must be noticed that our approach to the computation of the Casimir energy is still macroscopic.  
In order to fully describe its variation across the metal-superconductor transition, a microscopic description would be 
necessary. We do not have such a picture at the moment, but work is in progress in this direction. 

\begin{appendix}

\section{}

The generating functions are obtained by imposing the most general boundary conditions at each singular layer located at  
$d_0,\, d_0+d_1,\,d_0+d_1+d_2$ ... etc. (see Fig.\,\ref{fig:ncav}). These are obtained, as usual, by integrating the Maxwell 
equations
\beqs
\bfs{\nabla }\cdot \bfs{D}=\rho,~~&&~~ \bfs{\nabla}\times \bfs{E}+\frac{\partial\bfs{B}}{\partial t}=0, \\  \\
\bfs{\nabla }\cdot \bfs{B}=0,~~&&~~  \bfs{\nabla}\times \bfs{H}-\frac{\partial\bfs{D}}{\partial t}=\bfs{J}, \\ \\
\bfs{D}= \epsilon \bfs{E},~~&&~~  \bfs{B}=\mu \bfs{H},~~~(\epsilon=\epsilon_0,~\mu=\mu_0  \mbox{ in vacuum}),
\eeqs
across the discontinuity layers \cite{jack98},
\begin{eqnarray*}
(\bfs{D}_i-\bfs{D}_{i-1})\cdot \bf{n}=\sigma,~~ &&~~ (\bfs{B}_i-\bfs{B}_{i-1})\cdot \bf{n}=0, \\
\bf{n}\times (\bfs{E}_i-\bfs{E}_{i-1})=0, ~~&&~~ \bf{n}\times (\bfs{H}_i-\bfs{H}_{i-1})=\bf{J}, 
\end{eqnarray*}
where $\bf{n}=\hat{z}$ is the normal to the layers (parallel to the $z$-axis, going from the $i$-th to the $i+1$-th layer), $\sigma$ 
is the surface charge density, and $\bf{J}$ is the surface current density respectively (in principle they could be different at 
each layer, but we will not consider this situation).  By virtue of the translational invariance in the $(x,y)$ plane we can set 
$$
\bfs{E}={\bfs{f}}(z)\mathrm e^{\mathrm i(\bfs{k}_{||}\cdot\bfs{x}_{||}-\omega t)},~~~
\bfs{B}=\bfs{g}(z)\mathrm e^{\mathrm i(\bfs{k}_{||}\cdot\bfs{x}_{||}-\omega t)}.
$$
In the following, when discussing the plasma sheet model, we will consider the so called hydrodynamic model 
\cite{barton_2005,bordag_2005}, in which a continuous fluid with mass $m^*$ and charge $q^*$ is uniformly distributed in the layer 
with an overall-neutralizing background charge. The fluid displacement $\bfs{\xi}$ is purely tangential, 
$\bfs{\xi}\equiv (\xi_x,\xi_y)$ with surface charge and current densities related to the tangential component of the electric field 
$\bfs{E}_\parallel$ by
\begin{equation*}
\ddot{\boldsymbol{\xi}}=\frac{q^*}{m^*}{\bf{E}}_{\parallel}=: {q_0} {\bf{E}}_{\parallel},~~{\bf{J}}=
n_{2D}\, q^* \dot{\bfs{\xi}}=: \sigma_0 \dot{\bfs{\xi}},~~\sigma=-n_{2D} \, q^*{\bfs{\nabla}}_{\parallel}\cdot{\bfs{\xi}}=- \sigma_0{\bfs{\nabla}}_{\parallel}\cdot{\bfs{\xi}},
\end{equation*}
$n_{2D}$ being the two dimensional carrier density in the layer, and $q^*$ and $m^*$ being their charge and mass, respectively.
Under these assumptions, the most general boundary conditions at the $i=1,2,3 ...$-th boundaries are
\beq
\epsilon_i(\omega) f_i^z-\epsilon_{i-1}(\omega) f_i^z &=&-\frac{\sigma_0 q_0}{\omega^2} \frac{\partial f^z_i}{\partial z},\\
\frac{\partial f^z_i}{\partial z}-\frac{\partial f^z_{i-1}}{\partial z}&=&0, ~~~~~\mbox{for the TM modes, and} \nonumber \\
\nonumber \\
\frac{\partial g^z_i}{\partial z}-\frac{\partial g^z_{i-1}}{\partial z}&=&\Omega\; g^z_i,\\
g_i^z-g_{i-1}^z&=&0, ~~~~~\mbox{for the TE modes}, \nonumber
\eeq
with $\Omega={\mu_0\sigma_0 q_0}$.
With these boundary conditions, the generating functions for the $\rm TM$ and $\rm TE$ modes, respectively, can be written in 
terms of the auxiliary functions 
\begin{eqnarray*}
R_{\rm TM}^{i,j} &=& \frac{\epsilon_j(\mathrm i\zeta_l)K_i-\epsilon_i(\mathrm i\zeta_l)K_j - 
{2\frac{\Omega}{\zeta_l^2}} K_iK_j}{\epsilon_j(\mathrm i\zeta_l)K_i+\epsilon_i(\mathrm i\zeta_l)K_j + 
{2\frac{\Omega}{\zeta_l^2}} K_iK_j},~~~R_{\rm TE}^{i,j} = \frac{K_i-K_j +2 \Omega}{K_i+K_j +2 \Omega} \\
 \\
S_{\rm TM}^{i,j} &=&\frac{\epsilon_j(\mathrm i\zeta_l)K_i-\epsilon_i(\mathrm i\zeta_l)K_j + {2\frac{\Omega}{\zeta_l^2}} 
K_iK_j}{\epsilon_j(\mathrm i\zeta_l)K_i+\epsilon_i(\mathrm i\zeta_l)K_j + {2\frac{\Omega}{\zeta_l^2}} K_iK_j},~~~
S_{\rm TE}^{i,j} = \frac{K_i-K_j -2 \Omega}{K_i+K_j +2 \Omega} \\ \\
T_{\rm TM}^{i,j} &=&\frac{\epsilon_j(\mathrm i\zeta_l)K_i+\epsilon_i(\mathrm i\zeta_l)K_j - {2\frac{\Omega}{\zeta_l^2}} 
K_iK_j}{\epsilon_j(\mathrm i\zeta_l)K_i+\epsilon_i(\mathrm i\zeta_l)K_j + {2\frac{\Omega}{\zeta_l^2}} K_iK_j},
~~~T_{\rm TE}^{i,j} = \frac{K_i+K_j -2 \Omega}{K_i+K_j +2 \Omega} 
\end{eqnarray*}
with $K_i=\sqrt{k_\perp^2+\epsilon_i(\mathrm i\zeta_l) \zeta^2_l}$, and $k_\perp=(k_x,k_y)$. Standard dielectric boundary conditions 
(dbc) are recovered by imposing $\Omega=0$ and the plasma sheet boundary conditions (psbc) by requiring 
$\epsilon_j(\mathrm i\zeta_l)=1,~~\forall j$ (in this case $K_i=K_j$).
\begin{eqnarray*}
{E}^{ijk}  &=&  {\mathrm e^{-2 d_j K_j} \text{S}^{j,k} \text{R}^{i,j}}+1, \\
F^{ijk}  &=&{\mathrm e^{-2 d_j K_j} \text{R}^{i,j}\text{T}^{j,k}}+\text{R}^{j,k}, \\
G^{ijk} &=& {\mathrm e^{-2 d_j K_j} \text{S}^{j,k} \text{T}^{i,j}}+{\text{S}^{i,j}}, \\
H^{ijk}   &=&{\text{S}^{i,j} \text{R}^{j,k}}+{\mathrm e^{-2 d_j K_j} \text{T}^{i,j} \text{T}^{j,k}}.
\end{eqnarray*}
Considering henceforth all the cavities to be equal, we consider only the indices $\{ijk\}= \{012\}$, so that
\begin{eqnarray*}
I_1 &=& E^{012};~~I_2=F^{012}e^{-2 d_2  K_2}G^{012},\\
I_n &=&F^{012}e^{-2 d_2  K_2}\left(H^{012}e^{-2 d_2  K_2}\right)^{n-2}G^{012},~~\mbox{      for $n\geq3$}.
\end{eqnarray*}

Let us consider, for the case of the TM modes, three cavities. Letting $x_i=\sum_{n=0}^i d_n$, the
matching conditions give rise to the $12 \times 12$ 
matrix of coefficients
\begin{center}
\scalebox{.65}{
$
M=\left(
\begin{array}{cccccccccccc}
 -{\mathrm e}^{x_0 K_0}\epsilon_0 &{\mathrm e}^{-x_0 K_1} \epsilon_1 &{\mathrm e}^{x_0 K_1}  \epsilon_1 & 0 & 0 & 0 & 0 & 0 & 0 & 0 & 0 & 0  \\
 -{\mathrm e}^{-x_0 K_0} K_0 &{\mathrm e}^{-x_0 K_1}  K_1 & -{\mathrm e}^{x_0 K_1} K_1 & 0 & 0 & 0 & 0 & 0 & 0 & 0 & 0 & 0 \\
 0 & {\mathrm e}^{x_1 K_1} \epsilon_1 & {\mathrm e}^{-x_1 K_1} \epsilon_1 & -{\mathrm e}^{-x_1 K_2}
   \epsilon_2 & -{\mathrm e}^{x_1 K_2} \epsilon_2 & 0 & 0 & 0 & 0 & 0 & 0 & 0 \\
 0 & {\mathrm e}^{x_1 K_1} K_1 & -{\mathrm e}^{-x_1 K_1} K_1 & {\mathrm e}^{-x_1 K_2} K_2 & -{\mathrm e}^{x_1
   K_2} K_2 & 0 & 0 & 0 & 0 & 0 & 0 & 0 \\
 0 & 0 & 0 & {\mathrm e}^{-K_2 x_2} \epsilon_2 & {\mathrm e}^{K_2 x_2} \epsilon_2 & -{\mathrm e}^{K_3
   x_2} \epsilon_3 & -{\mathrm e}^{-K_3 x_2} \epsilon_3 & 0 & 0 & 0 & 0 & 0 \\
 0 & 0 & 0 & -{\mathrm e}^{-K_2 x_2} K_2 & {\mathrm e}^{K_2 x_2} K_2 & -{\mathrm e}^{K_3 x_2} K_3 &
   {\mathrm e}^{-K_3 x_2} K_3 & 0 & 0 & 0 & 0 & 0 \\
 0 & 0 & 0 & 0 & 0 & {\mathrm e}^{K_3 x_3} \epsilon_3 & {\mathrm e}^{-K_3 x_3} \epsilon_3 &
   -{\mathrm e}^{-K_4 x_3} \epsilon_4 & -{\mathrm e}^{K_4 x_3} \epsilon_4 & 0 & 0 & 0 \\
 0 & 0 & 0 & 0 & 0 & {\mathrm e}^{K_3 x_3} K_3 & -{\mathrm e}^{-K_3 x_3} K_3 & {\mathrm e}^{-K_4 x_3}
   K_4 & -{\mathrm e}^{K_4 x_3} K_4 & 0 & 0 & 0 \\
 0 & 0 & 0 & 0 & 0 & 0 & 0 & {\mathrm e}^{-K_4 x_4} \epsilon_4 & {\mathrm e}^{K_4 x_4}
   \epsilon_4 & -{\mathrm e}^{K_5 x_4} \epsilon_5 & -{\mathrm e}^{-K_5 x_4} \epsilon_5 & 0  \\
 0 & 0 & 0 & 0 & 0 & 0 & 0 & -{\mathrm e}^{-K_4 x_4} K_4 & {\mathrm e}^{K_4 x_4} K_4 &
   -{\mathrm e}^{K_5 x_4} K_5 & {\mathrm e}^{-K_5 x_4} K_5 & 0 \\
 0 & 0 & 0 & 0 & 0 & 0 & 0 & 0 & 0 & {\mathrm e}^{K_5 x_5} \epsilon_5 & {\mathrm e}^{-K_5 x_5}
   \epsilon_5 & -{\mathrm e}^{-K_6 x_5} \epsilon_6 \\
 0 & 0 & 0 & 0 & 0 & 0 & 0 & 0 & 0 & {\mathrm e}^{K_5 x_5} K_5 & -{\mathrm e}^{-K_5 x_5} K_5 &
   {\mathrm e}^{-K_6 x_5} K_6 ,\\
\end{array}
\right).
$
}
\end{center}
\bigskip
Computing the determinant of the minors of dimensions $4,\,8$, and $12$, we obtain the energy of one, two, and three cavities, 
respectively. After regularization, for the single cavity $(012)$ in Fig.\,1, we find
\begin{equation}
\Delta_{1} = {E}^{012}=I_1.
\end{equation}
For two cavities $(012-234)$, we find
\beq
\Delta_{2} &=& {E}^{012}{E}^{234}+\mathrm e^{-2(d_2k_2)}F^{012} G^{234}=:(I_1)^2 +I_{2}, \mbox{ ~~~   and}   \nonumber\\ 
\log\Delta_{2} &=&\log(I_1^2) +\log\left(1+\frac{I_{2}}
{I_1^2} \right). 
\eeq
Finally, for the three cavities, we find
\begin{eqnarray}
\Delta_{3} &=& E^{012}E^{234}E^{456}+
\mathrm e^{-2(d_2k_2+d_4k_4)}F^{012} H^{234}G^{456}   \nonumber \\
&& +\mathrm e^{-2 d_2k_2}E^{456}F^{012} G^{234}+ 
\mathrm e^{-2 d_4k_4}E^{012}F^{234} G^{456} \nonumber   \\
&&=:I_1^3+I_{3}+I_1I_{2}+I_1I_{2},  \\
\log\Delta_{3} &=&\log\left(I_1^3\right) +
\log\left(1+\frac{2I_1I_{2}}{I_1^3}\right)+
\log\left(1+\frac{I_{3}}{I_1^3+2I_1I_2} \right).  
\end{eqnarray}
When $d_2\rightarrow\infty$, $I_{2}\rightarrow0$ and
$$
\log{ \Delta_{2}} = \log{I_1^2}=2\log{I_1}.
$$ 
Thus, when the two cavities are far away, their energy is simply the sum of the individual contributions and  
$I^{(2)}$ can be seen as the energy due to the coupling of the two cavities $(012)-(234)$.
For the three cavities $(012-234-456)$, the formulas are written so as to make evident the 
contribution to the energy resulting from the sum of the energies of the single cavities, with respect to 
the one coming from the coupling of the two possible pairs of cavities $(012-234),~(234-456)$, 
and the one coming from the coupling of the three, $I^{(3)}$. 

\end{appendix}

\end{document}